\documentclass[showpacs]{revtex4}
\usepackage[caption=false]{subfig}
\usepackage{epsfig}
\usepackage{color}
\def\beq{\begin{equation}}
\def\enq{\end{equation}}
\def\beqa{\begin{eqnarray}}
\def\enqa{\end{eqnarray}}
\def\nnb{\nonumber}

\def\MeV{\nobreak\,\mbox{MeV}}
\def\GeV{\nobreak\,\mbox{GeV}}

\def\qslash{q\kern-.5em\slash}

\def\qq{\lag\bar{q}q\rag}

\def\Gd{\lag g^2G^2\rag}
\def\G3{\lag g^3G^3\rag}

\def\GG{\lag G^2 \rag}
\def\qGq{\lag \bar{q}Gq \rag}
\def\GGG{\lag G^3 \rag}

\def\al{\alpha}

\def\lb{\label}
\def\nn{\nonumber}

\newcommand{\rag}{\rangle}
\newcommand{\lag}{\langle}

\begin{document}

\title{\sc QCD sum rule study for a possible charmed pentaquark $\Theta_c(3250)$ }
\author{Raphael M. Albuquerque}
\email{raphael@ift.unesp.br}
\affiliation{Institute for Theoretical Physics, S\~ao Paulo State University (IFT-UNESP)\\
R. Dr. Bento Teobaldo Ferraz, 271 - Bloco II, 01140-070 S\~ao Paulo, SP - Brazil}

\author{Su Houng Lee}
\email{suhoung@phya.yonsei.ac.kr}
\affiliation{Institute of Physics and Applied Physics, Yonsei University,
Seoul 120-749, Korea}

\author{Marina Nielsen}
\email{mnielsen@if.usp.br}
\affiliation{Instituto de F\'{\i}sica, Universidade de S\~{a}o Paulo,
C.P. 66318, 05389-970 S\~{a}o Paulo, SP, Brazil}

\begin{abstract}
We use QCD  sum rules to study the possible existence of a $\Theta_c(3250)$
charmed pentaquark.
We consider the contributions of condensates up to dimension-10 and
work at leading order in $\alpha_s$. We obtain $m_{\Theta_c}=(3.21\pm0.13)~\GeV$,
compatible with the mass of the structure seen by BaBar Collaboration
in the decay channel $B^-\to\bar{p} \:\Sigma_c^{++} \:\pi^-\pi^-$.  The proposed
state is compatible with a previous proposed pentaquark state in the
anti-charmed sector.
\end{abstract}


\maketitle

\section{Introduction}
Recently, the BaBar Collaboration has reported \cite{Lees:2012kc} the observation 
of unexplained structures in the $B^-\to\bar{p} \:\Sigma_c^{++}\pi^-\pi^-$ decay channel. 
In particular, they observed three enhancements in the $\Sigma^{++}_c \:\pi^-\pi^-$
invariant mass distribution at $3.25~\GeV$, $3.80~\GeV$ and $4.20~\GeV$
\cite{Lees:2012kc}. We shall refer to these signals $\Theta_c(3250)$,
$\Theta_c(3800)$ and $\Theta_c(4200)$, respectively.
There are already theoretical calculations interpreting the $\Theta_c(3250)$ enhancement
as a possible $D_0^*(2400) \:N$ molecular state \cite{He:2012zd,Zhang:2012xx}. 
In this note we follow a different approach, and we use the QCD sum rules (QCDSR)
\cite{svz,rry,SNB} to try to interpret $\Theta_c(3250)$ enhancement as a charmed
pentaquark.

There are already some calculations for charmed pentaquarks.
Based on simple theoretical considerations, Diakonov has predicted the masses of
the exotic anti-decapenta-plet of charmed pentaquarks \cite{Diakonov:2010tf}.
In his model, the lightest members of this multiplet are explicitly exotic doublets,
$cuud\bar{s}$ and $cudd\bar{s}$, with mass about 2.42 GeV. The crypto-exotic
$cudd\bar{u}$ pentaquark should have a mass around 140 MeV heavier. Since 
the accuracy of this prediction is $\sim150~\MeV$, Diakonov's prediction for the
mass of the $cudd\bar{u}$ pentaquark is $\sim50~\MeV$ smaller than the observed
enhancement. Using the Skyrme soliton model Wu and Ma have studied the exotic
pentaquark states with charm and anti-charm \cite{Wu:2004wg}. In their approach, they
obtained a mass around 2.70 GeV for both $cudd\bar{u}$ and $uudd\bar{c}$ states.

The first QCDSR calculation for a possible anti-charmed pentaquark was done in
Ref.~\cite{Kim:2004pu}. The authors have found a mass around $3.10 \GeV$, supposing that 
the anti-charmed pentaquark can be described by a current with two-light diquarks and 
one anti-charm quark. Since for a charmed $cudd\bar{u}$ pentaquark one needs a light 
diquark, a heavy-light diquark and a light antiquark to describe it, and since light diquarks 
are supposed to be very bound states \cite{Shuryak:2005pk} and heavy-light diquarks less 
bound \cite{Lee:2009rt}, we expect the mass of the charmed pentaquark to be 
bigger than the mass of the anti-charmed pentaquark and, therefore, compatible with the
observed $\Theta_c(3250)$ enhancement.

\section{Two-Point Correlation Function}
A possible current describing a charmed neutral pentaquark with quark content
$[cudd\bar{u}]$, which we call $\Theta_{1c}$, is given by:
\beq
\eta_{1c}=\varepsilon^{abc}(\varepsilon^{aef} \:{\bf u}_e^TC\gamma_5 {\bf d}_f) \:
(\varepsilon^{bgh} \:{\bf c}_g^TC \gamma_5 {\bf d}_h) \:C \,\gamma_5 \:\bar{{\bf u}}_c^T\;,
\label{field}
\enq
where $a,~b,...$  are color indices, $C$ is the charge conjugation matrix and in bold letters 
are the respective quark fields. We have considered two scalar diquarks since they are 
supposed to be 
more bound than the pseudoscalars \cite{Shuryak:2005pk}. However, since the study presented in
\cite{Shuryak:2005pk} is related with the light diquarks, one could also have a
current describing another pentaquark, $\Theta_{2c}$, with a scalar light-diquark and
a pseudoscalar heavy-light-diquark as follows:
\beq
\eta_{2c}=\varepsilon^{abc}(\varepsilon^{aef} \:{\bf u}_e^TC\gamma_5 {\bf d}_f) \:
(\varepsilon^{bgh} \:{\bf c}_g^TC {\bf d}_h) \:C \:\bar{{\bf u}}_c^T\;,
\label{field2}
\enq
like the current used in \cite{Kim:2004pu} for $\Theta_{\bar{c}}$.
The sum rule for both currents (\ref{field}) and (\ref{field2}) is constructed
from the two-point correlation
\beq
\Pi(q)=i\int d^4x ~e^{iq \cdot x}\lag 0
|T[\eta_{c}(x)\bar{\eta}_{c}(0)]|0\rag=
\Pi_1(q^2) ~+~ \qslash \:\Pi_2(q^2) ~,
\lb{2po}
\enq
where $\Pi_1$ and $\Pi_2$ are two invariant independent functions.
In the phenomenological side, we parametrize the spectral function using the
standard duality ansatz: ``one resonance"+ ``QCD continuum".
The QCD continuum starts from a threshold $s_0$ and comes from
the discontinuity of the QCD diagrams. Transferring its contribution to the QCD side
of the sum rule, one obtains the Borel/Laplace sum rules:
\beqa
|\lambda_{_{\Theta_c}}|^2 \:m_{_{\Theta_c}}~e^{-m_{_{\Theta_c}}^2 / M_B^2} &=&
\int_{m_c^2}^{s_0}ds~e^{-s/M_B^2}~\rho_1(s)~,\nnb\\
|\lambda_{_{\Theta_c}}|^2 ~e^{-m_{_{\Theta_c}}^2 / M_B^2} &=&
\int_{m_c^2}^{s_0}ds ~e^{-s/M_B^2}~\rho_2(s)~,
\lb{srm}
\enqa
where $\rho_i = {1\over\pi}{\rm Im}\,\Pi_i(s)$ are the spectral
densities whose expressions are given in the Appendix. In Eq.~(\ref{srm}),
$\lambda_{_{\Theta_c}}$ and $m_{_{\Theta_c}}$ are the pentaquark residue and mass,
respectively; $M_B^2$ is the sum rule variable.
One can estimate the pentaquark mass from the following ratios
\beqa
{\cal R}_i &=& {\int_{m_c^2}^{s_0}ds~s~
e^{-s/M_B^2}~\rho_{i}(s)\over \int_{m_c^2}^{s_0}ds~
e^{-s/M_B^2}~\rho_{i}(s)}~,~~~~~i=1,2~,\nnb\\
{\cal R}_{12} &=& {\int_{m_c^2}^{s_0}ds~
e^{-s/M_B^2}~\rho_{1}(s)\over \int_{m_c^2}^{s_0}ds~
e^{-s/M_B^2}~\rho_{2}(s)}~,
\label{eq:ratio}
\enqa
where at the $M_B^2$-stability point, we have
\beq
m_{_{\Theta_c}} ~\simeq~ \sqrt{{\cal R}_i} ~\simeq~
{\cal R}_{12}~.
\label{eqmass}
\enq

\section{Numerical Results}
For a consistent comparison with the results obtained for other pentaquark
states using the QCDSR approach, we have considered the same values
used for the heavy quark mass and condensates as in Ref.~\cite{SNB,narpdg},
listed in Table \ref{TabParam}.
It is worth mentioning that, for both currents $\eta_{1c}$ and $\eta_{2c}$,
we have found a substantial $M_B^2$-instability in the ${\cal R}_{12}$
sum rule evaluation. Therefore, in this work, we will only consider the results 
from ${\cal R}_{i}$.

{\small
\begin{table}[b]
\setlength{\tabcolsep}{1.25pc}
\caption{QCD input parameters.}
\begin{tabular}{ll}
&\\
\hline
Parameters&Values\\
\hline
$m_c$ & $(1.23 - 1.47) \GeV$ \\
$\qq$ & $-(0.23 \pm 0.03)^3\GeV^3$\\
$\lag g_s^2 G^2 \rag$ & $(0.88 \pm 0.25) ~\GeV^4$\\
$\lag g_s^3 G^3 \rag$ & $(0.58 \pm 0.18)~\GeV^6$\\
$m_0^2 \equiv \qGq / \qq$ & $(0.8 \pm 0.1) \GeV^2$\\
\hline
\end{tabular}
\label{TabParam}
\end{table}}

\begin{figure}[tp]
    \subfloat[]{\includegraphics[width=7.5cm]{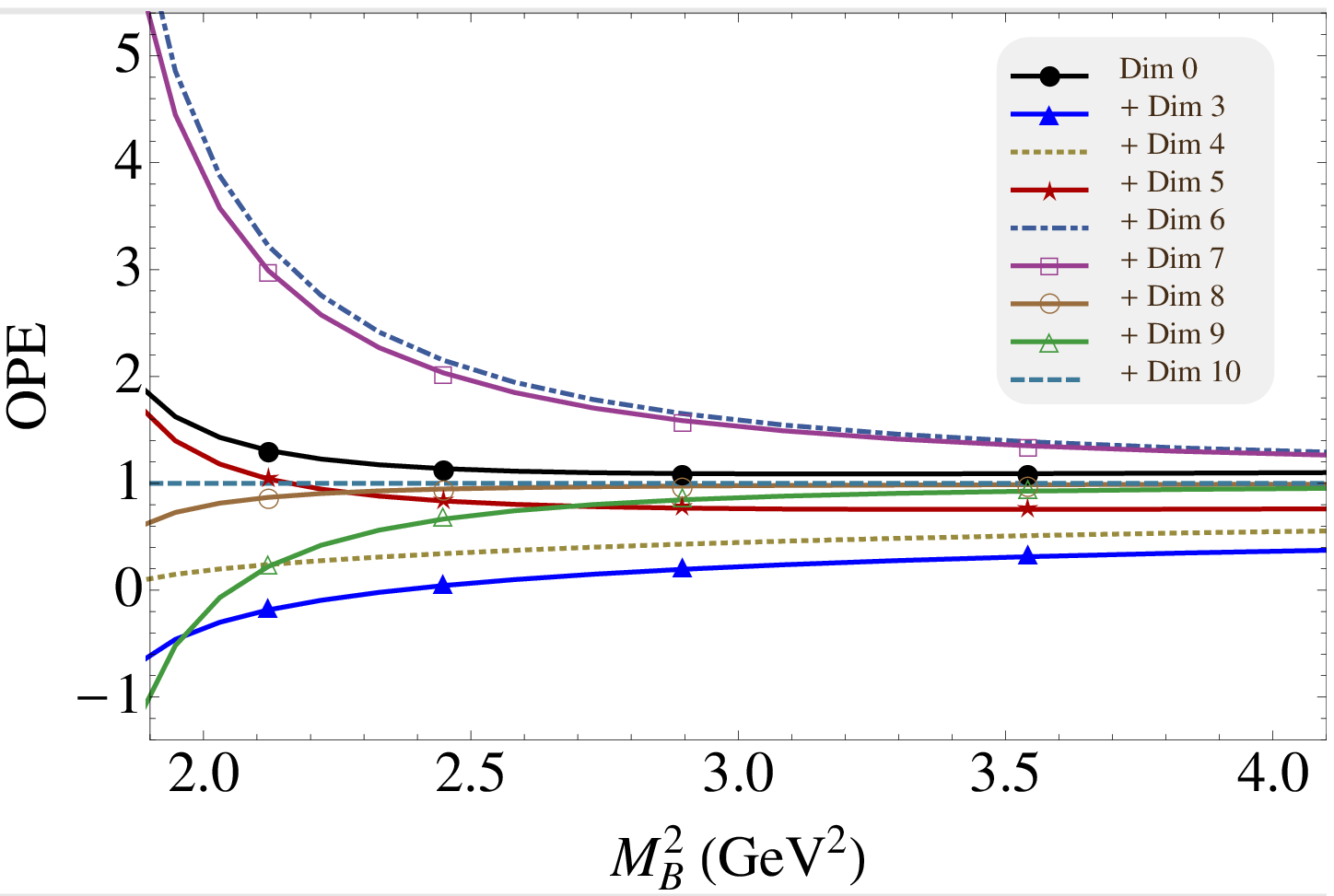}} \hspace{0.8cm}
    \subfloat[]{\includegraphics[width=7.5cm]{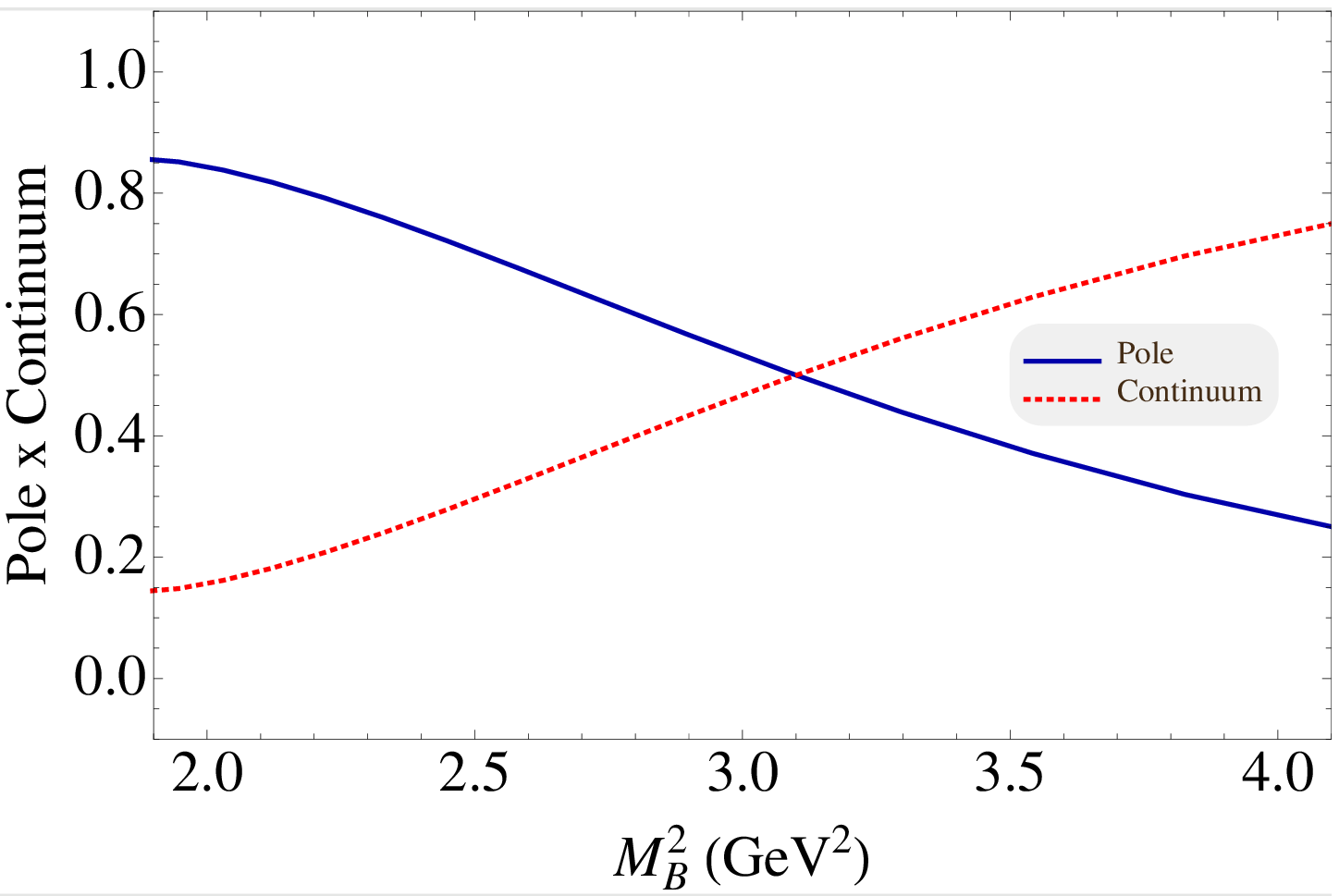}}\\
    \subfloat[]{\includegraphics[width=10cm]{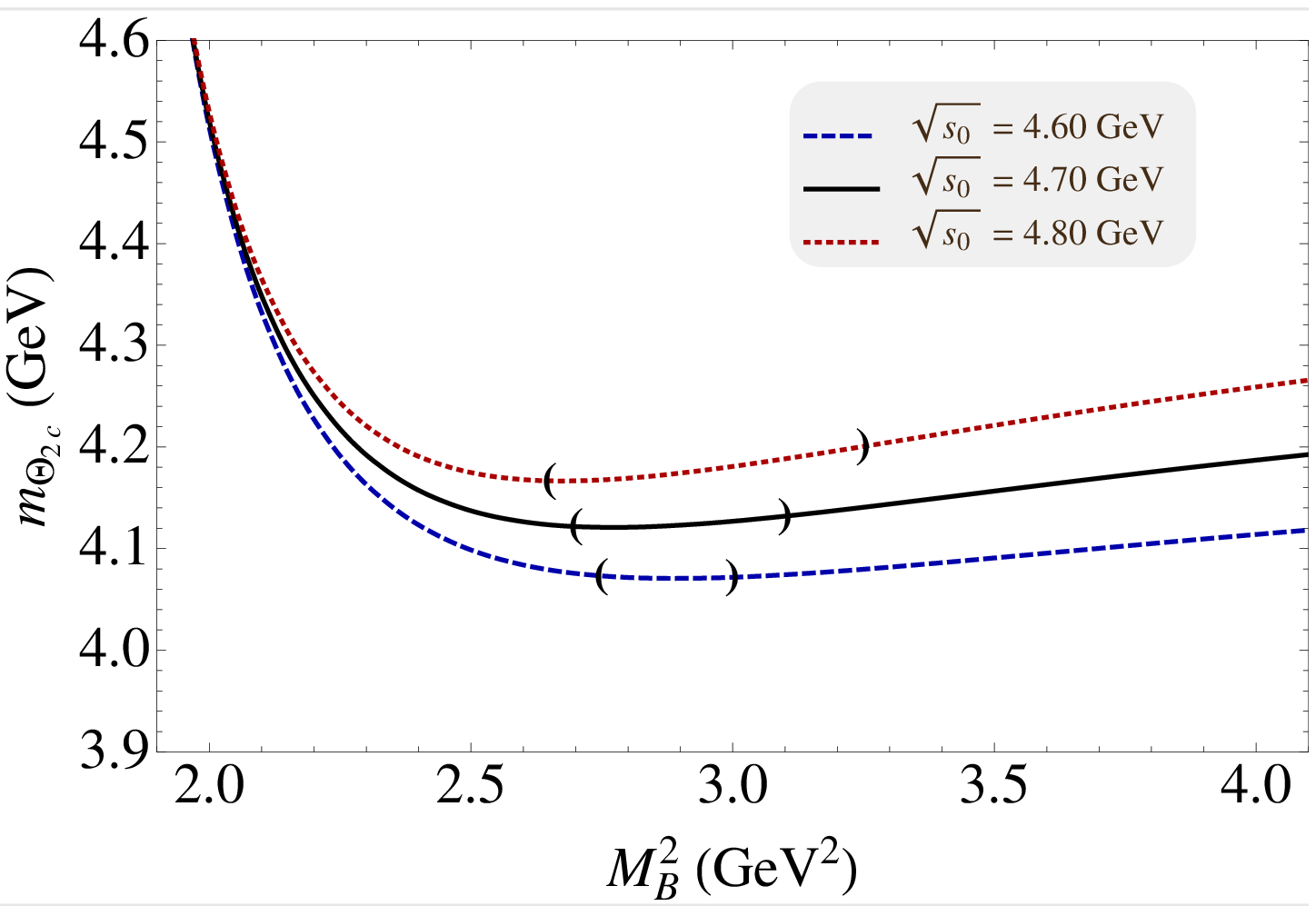}}
\caption{\footnotesize ${\cal R}_2$ sum rule analysis using the
pentaquark current $\eta_{2c}$. We have considered contributions up to
dimension-10 in the OPE, using $m_c=1.23\GeV$.
{\bf a)} OPE convergence in the region $(2.0 \leq M_B^2 \leq 4.0)~\GeV^{-2}$ for
$\sqrt{s_0} = 4.70 \GeV$. We plot the relative contributions starting with the
perturbative contribution and each other line represents the relative contribution
after adding of one dimension in the OPE expansion.
{\bf b)} The relative pole and continuum contributions for $\sqrt{s_0} = 4.70 \GeV$.
{\bf c)} The mass as a function of the sum rule parameter $M_B^2$, for different
values of $\sqrt{s_0}$. For each line, the region bounded by parenthesis indicates a
valid Borel window.}
\label{FigJ0}
\end{figure}
%

\subsection{$\Theta_{2c}$ Pentaquark State}
We start our analysis with the current $\eta_{2c}$. As mentioned above, we
calculate the mass related to this current using only the results from the
${\cal R}_1$ and ${\cal R}_2$ sum rules.

Considering the ${\cal R}_2$ sum rule, we show in Fig.~\ref{FigJ0}a) the relative contributions
of the terms in the OPE, for $\sqrt{s_0} = 4.70 \GeV$. From this figure, we see that the
contribution of the dimension-10 condensate is smaller than $20\%$ of the total contribution
for values of $M_B^2\geq 2.7 \GeV^2$, which indicates the starting point for a good
OPE convergence.
In Fig.~\ref{FigJ0}b), we also see that the pole contribution is bigger than the
continuum contribution only for values $M_B^2\leq 3.1 \GeV^2$. Therefore, we can
fix the Borel window as: $(2.7 \leq M_B^2 \leq 3.1) \GeV^2$.
From Eq.~(\ref{eqmass}), we can estimate the ground state mass, which is shown, as
a function of $M_B^2$, in Fig.~\ref{FigJ0}c). We conclude that there is a very good
$M_B^2$-stability in the determined Borel window, which is indicated through the
parenthesis.

Varying the value of the continuum threshold in the range $\sqrt{s_0} = 4.70 \pm 0.10 \GeV$,
and other parameters as indicated in Table \ref{TabParam}, we get
\begin{equation}
  m_{\Theta_{2c}} = 4.15 \pm 0.11 \GeV ~.
  \label{J0mass}
\end{equation}

This mass is surprisingly compatible with one of the unexplained structures observed
by BaBar Collaboration \cite{Lees:2012kc} at $4.2 \GeV$.
Therefore, from a sum rule point of view, such a $\Theta_{2c}$ pentaquark state with a
internal structure composed by a scalar light-diquark and a pseudoscalar heavy-light-diquark
could be a good candidate to explain the $\Theta_c(4200)$ enhancement.

\begin{figure}[t]
\centerline{\epsfig{figure=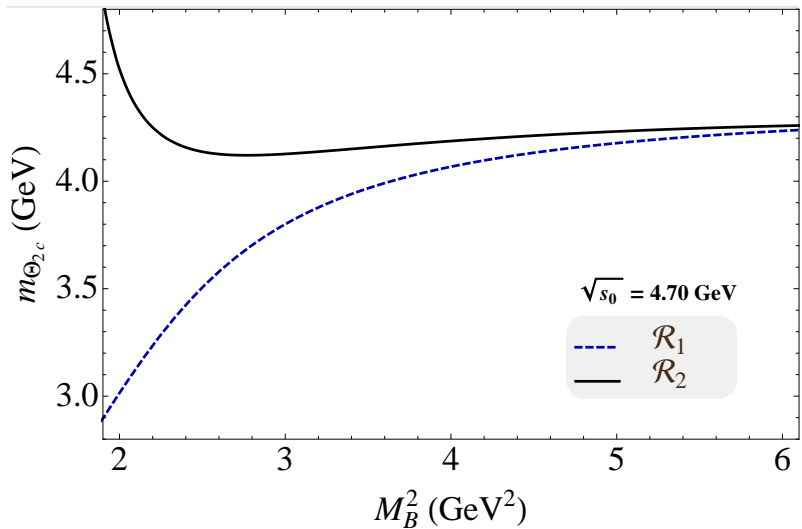,height=70mm}}
\caption{The comparison between the mass results evaluated with the
${\cal R}_1$ (dashed line) and ${\cal R}_2$
(solid line) sum rules, in the region $(2.0 \leq M_B^2 \leq 6.0) \GeV^2$ for
$\sqrt{s_0} = 4.70$ GeV.}
\label{FigJ0rmXrq}
\end{figure}

For completeness, we evaluate the ${\cal R}_1$ sum rule for the current $\eta_{2c}$. We
would naively expect to obtain a mass in accordance with Eq.~(\ref{J0mass}). The
comparison between the two sum rules is shown in Fig.~\ref{FigJ0rmXrq}, considering the
Borel range $(2.0 \leq M_B^2 \leq 6.0) \GeV^2$ and $\sqrt{s_0} = 4.70 \GeV$.
As one can see, the ${\cal R}_2$ sum rule presents a better $M_B^2$-stability than
${\cal R}_1$. Besides, the Borel window for the ${\cal R}_1$ sum rule lies on the range
$(2.5 \leq M_B^2 \leq 3.4) \GeV^2$ which does not contain $M_B^2$-stability.
Thus, we conclude that the results extracted from the ${\cal R}_1$ can be ruled out, while the
${\cal R}_2$ provides a more reliable sum rule calculation and the mass found in
Eq.~(\ref{J0mass}) must be settled as the optimized estimation for the $\Theta_{2c}$
pentaquark mass.

\subsection{$\Theta_{1c}$ Pentaquark State}
\begin{figure}[tp]
    \subfloat[]{\includegraphics[width=7.5cm]{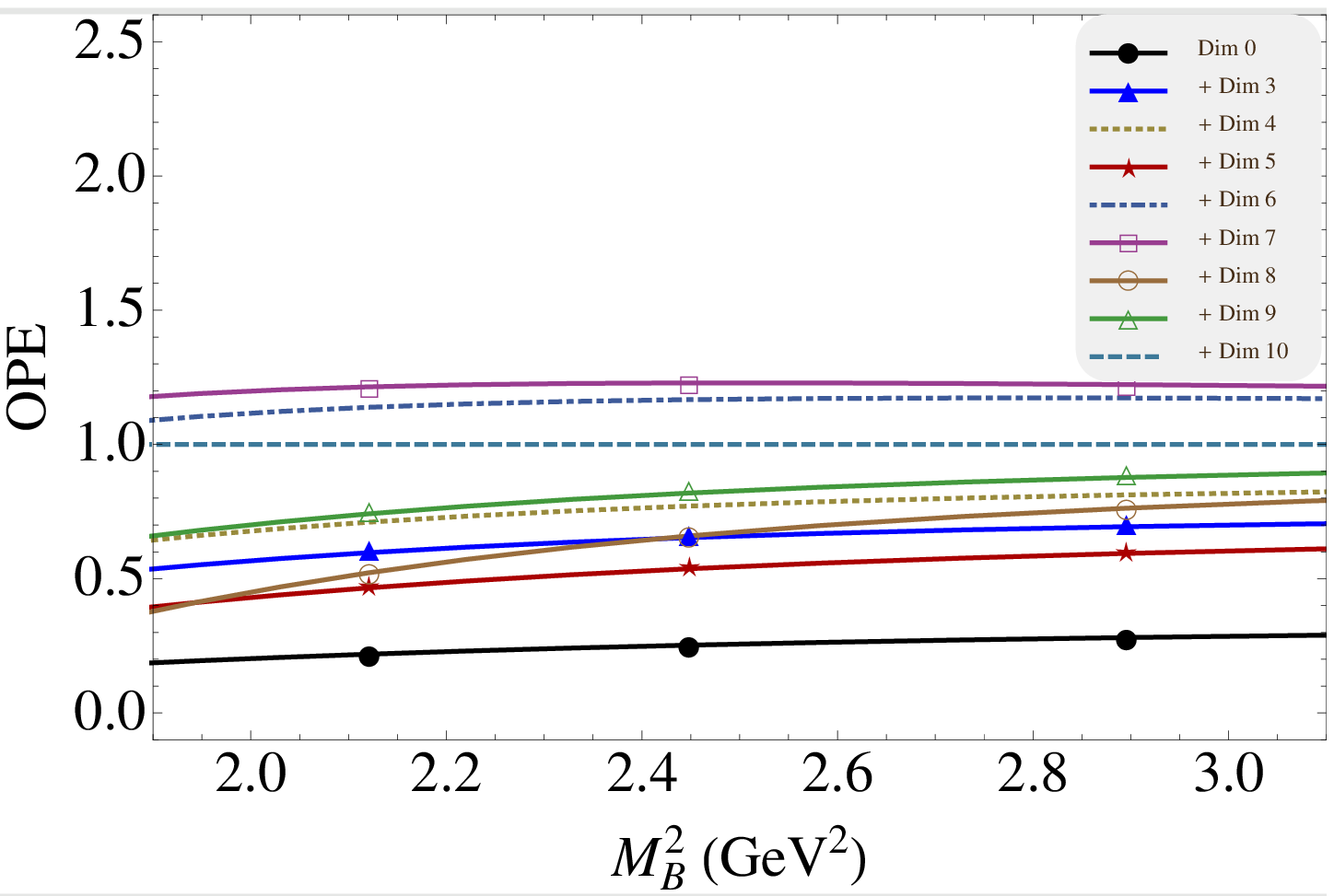}} \hspace{0.8cm}
    \subfloat[]{\includegraphics[width=7.5cm]{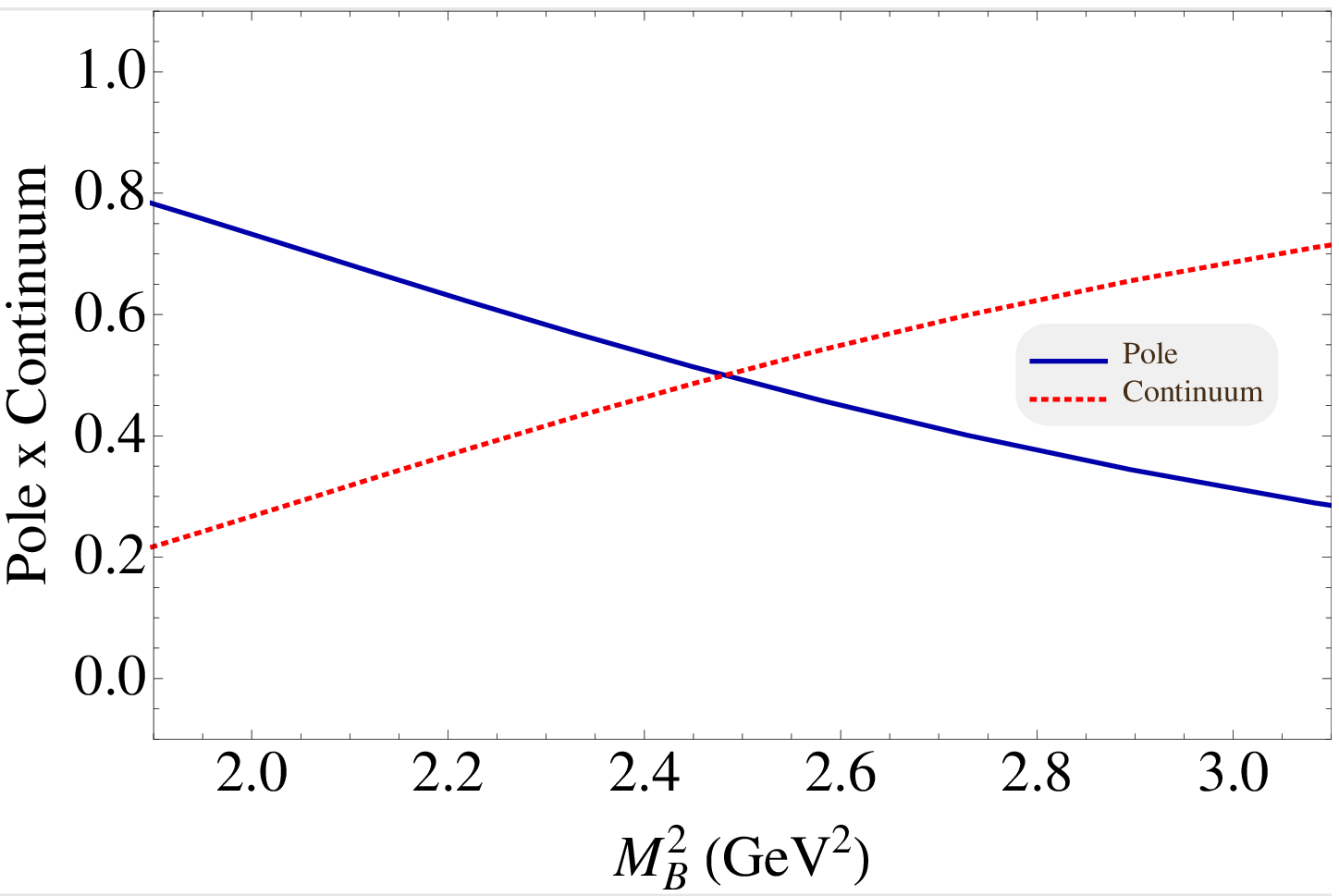}}\\
    \subfloat[]{\includegraphics[width=10cm]{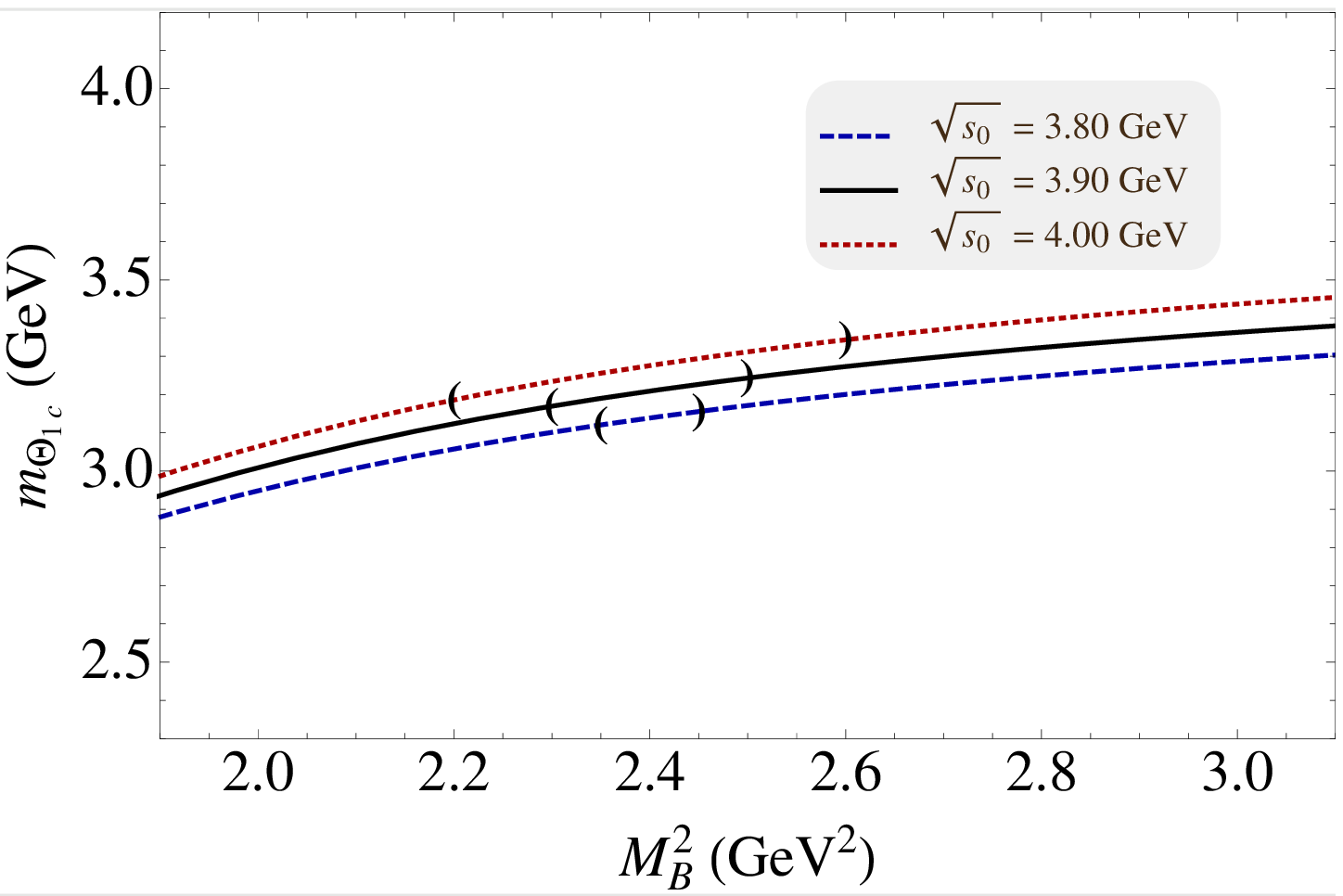}}
\caption{\footnotesize ${\cal R}_2$ sum rule analysis using the
pentaquark current $\eta_{\Theta_{1c}}$. We have considered contributions up to
dimension-10 in the OPE, using $m_c=1.23\GeV$.
{\bf a)} OPE convergence in the region $(1.8 \leq M_B^2 \leq 3.2)~\GeV^{-2}$ for
$\sqrt{s_0} = 3.90 \GeV$. We plot the relative contributions starting with the
perturbative contribution and each other line represents the relative contribution
after adding of one dimension in the OPE expansion.
{\bf b)} The relative pole and continuum contributions for $\sqrt{s_0} = 3.90 \GeV$.
{\bf c)} The mass as a function of the sum rule parameter $M_B^2$, for different
values of $\sqrt{s_0}$. For each line, the region bounded by parenthesis indicates a
valid Borel window.}
\label{FigJ1}
\end{figure}

In the case of the current $\eta_{1c}$, we also retain only the results from the
${\cal R}_2$ sum rule, according to the previous analysis. The results for the
pole dominance and OPE convergence are shown in Fig.~\ref{FigJ1} a) and b),
respectively. From these figures, we can fix the Borel window as:
$(2.3 \leq M_B^2 \leq 2.5) \GeV^2$, for $\sqrt{s_0} = 3.90 \GeV$. As one can see,
from the Fig.~\ref{FigJ1} c), we obtain $M_B^2$-stability only for a narrow Borel
window. However, it is still possible to extract reliable results from this sum rule.
Varying the continuum threshold in the range $\sqrt{s_0} = 3.90 \pm 0.10 \GeV$,
and the other parameters as indicated in Table \ref{TabParam}, we get
\begin{eqnarray}
  m_{\Theta_{1c}} &=& 3.21 \pm 0.13 \GeV ~.
  \label{J1mass}
\end{eqnarray}
This value for the mass is compatible with the first signal observed in
Ref.~\cite{Lees:2012kc} at $3.25 \GeV$. Therefore, we conclude that the $\Theta_c(3250)$
state also can be described by a pentaquark containing two scalar diquarks
in its internal structure. It is very interesting to notice that we get a smaller mass
with the current with two scalar diquarks, when compared with the current with
one scalar and one pseudoscalar diquarks. Although we have one light and one
light-heavy diquarks, our results follow the phenomenology obtained by Shuryak
\cite{Shuryak:2005pk} for the light diquarks.

It is interesting to compare our results with the result in Ref.~\cite{Zhang:2012xx},
where the author evaluates the sum rule for the $D^\ast_0(2400) \:N$ molecule, since
such a molecular current can be rewritten in terms of a sum over pentaquark type
currents, by using  Fierz transformations \cite{nfl}.
Indeed, the result found in Ref.~\cite{Zhang:2012xx} is in agreement with
our result in  Eq.~(\ref{J1mass}), which was obtained with the current
in Eq.~(\ref{field}).
However, there are some points in the analysis done in Ref.~\cite{Zhang:2012xx} that
deserve consideration. In particular, to obtain a mass compatible with the
3.25 GeV enhancement observed by BaBar, the author of
Ref.~\cite{Zhang:2012xx}, had to  release the criteria of pole dominance and the usual
 good OPE convergence. In doing so, the analysis inevitably led to
a misleading definition of the Borel window, fixed as $(2.0 \leq M_B^2 \leq 3.0) \GeV^2$
 for the $D^\ast_0(2400) \:N$ molecule. Besides, one can see that there is also no
$M_B^2$-stability in such Borel window. Therefore, we believe that if the author of
Ref.~\cite{Zhang:2012xx} had imposed pole dominance, good OPE convergence and Borel
stability in his analysis he would have obtained a bigger value for the mass of the
$D^\ast_0(2400) \:N$ current.

\section{Conclusions}
In conclusion, we have presented a QCDSR calculation for the two-point function of two
possible pentaquark states, whose internal structure is composed of two scalar
diquarks, for $\Theta_{1c}$, and a scalar light-diquark plus a pseudoscalar
heavy-light-diquark, for $\Theta_{2c}$. As expected from phenomenology
\cite{Shuryak:2005pk}, we get a smaller mass with the current $\eta_{1c}$ containing 
two scalar diquarks, in comparison with the current $\eta_{2c}$ containing one scalar and 
one pseudoscalar diquarks. 
Also, we get a bigger mass for the $\Theta_{2c}$ pentaquark state when comparing 
with the one studied in Ref. \cite{Kim:2004pu}, where the authors considered for the 
$\Theta_{\bar{c}}$ state a current with two-light diquarks and one anti-charm quark. 
Indeed, this result is in agreement with the expectation that heavy-light 
diquarks are less bound than light diquarks \cite{Lee:2009rt}.
Our findings strongly suggest that at least two enhancements observed by BaBar
Collaboration, with a peak at $3.25 \GeV$ and $4.20 \GeV$,
decaying into $\Sigma^{++}_c \:\pi^-\:\pi^- $, could be understood as being such
pentaquarks.

\section*{Acknowledgements}
We would like to thank APCTP for sponsoring the workshop on 'Hadron Physics
at RHIC'.  The discussions during the workshop have led the authors to
collaborate on this subject.
This work has been partly supported by FAPESP and CNPq-Brazil,
by the Korea Research Foundation KRF-2011-0020333 and KRF-2011-0030621 
and by the German BMBF grant 06BO108I.

\appendix

\section{Spectral Densities}\label{App1}

The spectral densities expressions for the charmed neutral pentaquarks, $\Theta_{1c}$ and
$\Theta_{2c}$, described by the currents in Eq.~(\ref{field}) and (\ref{field2})
respectively, have been calculated up to dimension-10 condensates, at leading order in 
$\alpha_s$. To keep the heavy quark mass finite, we use the momentum-space expression 
for the heavy quark propagator.
We calculate the light quark part of the correlation function in the coordinate-space,
and we use the Schwinger parameters to evaluate the heavy quark part of the correlator.
To evaluate the $d^4x$ integration in Eq.~(\ref{2po}), we use again the Schwinger
parameters, after a Wick rotation. Finally we get integrals in the Schwinger parameters.
The result of these integrals are given in terms of logarithmic functions, from where we
extract the spectral densities and the limits of the integration. The same technique can
be used to evaluate the condensate contributions.

For the $\qslash$-structure of the correlation function (\ref{2po}), we get:
\beqa\label{eq:pert}
\rho_2^{pert}(s) &=& -{1\over 5^2 \cdot 3 \cdot 2^{15} \:\pi^8} \int\limits_0^\Lambda
\!\!d\al \:{\al^5 \:{\cal H}_\al^5 \over (1-\alpha)^4} ,
\nn\\
\rho_2^{\qq}(s)&=&(-1)^{j+1}{m_c\qq \over 3^2 \cdot 2^{11} \:\pi^6}\int\limits_{0}^{\Lambda}
\!\!d\al \:{\al^4 \:{\cal H}_\al^3 \over (1-\al)^3} ,
\nn\\
\rho_2^{\lag G^2\rag}(s)&=&-{\Gd\over 5\cdot 3^2 \cdot 2^{21} \:\pi^8}\int\limits_{0}^{\Lambda}
\!\!d\al \:{\al^3 \:{\cal H}_\al^2 \over (1-\alpha)^4} \Big[ 32 m_c^2 \al^2 + 5{\cal H}_\al (1-\al) (52-33\al) \Big],
\nn\\
\rho_2^{\qGq}(s)&=& (-1)^{j+1}{m_c \qGq \over 3 \cdot 2^{15} \:\pi^6}\int\limits_{0}^{\Lambda}
\!\!d\al \:{\al^3 \:{\cal H}_\al^2\over(1-\alpha)^3} (19-23\al),
\nn\\
\rho_2^{\qq^2}(s) &=& {\qq^2 \over 3\cdot 2^7 \:\pi^4}\int\limits_{0}^{\Lambda}
\!\!d\al \:{\al^2 \:{\cal H}_\al^2 \over 1-\alpha} ,
\nn\\
\rho_2^{\GGG}(s) &=& -{\G3\over 5\cdot 3^2 \cdot 2^{20} \:\pi^8}\int\limits_{0}^{\Lambda}
\!\!d\al \:{\al^4\:{\cal H}_\al \over(1-\alpha)^4} \Big[ 4m_c^2(95-91\al) + {\cal H}_\al(285 - 281\al) \Big],
\nn\\
\rho_2^{\qq \GG}(s) &=& (-1)^{j+1} {m_c \qq \Gd \over 3^3 \cdot 2^{15} \:\pi^6}\int\limits_{0}^{\Lambda}
\!\!d\al \:{\al^2 \over(1-\alpha)^3} \Big[ 4m_c^2 \al^2 + 3{\cal H}_\al \Big(49 - \al(119-74\al) \Big) \Big],
\nn\\
\rho_2^{\qq \qGq}(s) &=& {\qq \qGq \over 3 \cdot 2^{12} \:\pi^4} \int\limits_{0}^{\Lambda}
\!\!d\al \:{\al \:{\cal H}_\al\over(1-\al)} (70 - 73\al),
\nn\\
\rho_2^{\qq^3}(s) &=& (-1)^j {m_c \qq^3 \over 3^2 \cdot 2^{3} \:\pi^2} \int\limits_{0}^{\Lambda}
\!\!d\al \:\al,
\nn\\
\rho_2^{\GG \qGq}(s) &=& (-1)^{j+1}{m_c \Gd \qGq \over 3^3 \cdot 2^{17} \:\pi^6} \Bigg\{
  \int\limits_{0}^{\Lambda}\!\!d\al \:{3\al \over (1 \!-\! \al)^2} \Big( 39 \!-\! \al(89 \!-\! 66\al) \Big) -
  \int\limits_{0}^{1}\!\!d\al \:{16m_c^2 \al^3 \over (1-\al)^3} \delta \!\left( \!s \!-\! {m_c^2 \over 1\!-\!\al}\! \right)
  \Bigg\},
\nn\\
\rho_2^{\qq \GGG}(s) &=& (-1)^{j+1} {m_c \qq \G3 \over 3^3 \cdot 2^{14} \:\pi^6} \Bigg\{
  \int\limits_{0}^{\Lambda}\!\!d\al \: {3\al^3 \over (1-\al)^3} (9-8\al) -
  \int\limits_{0}^{1}\!\!d\al \:{m_c^2\al^3 \over (1-\al)^4}(6-5\al) \delta \!\left( \!s \!-\! {m_c^2 \over 1\!-\!\al} \right)
  \Bigg\},
\nn\\
\rho_2^{\qGq^2}(s) &=& {\qGq^2 \over 3^2 \cdot 2^{13} \:\pi^4}
  \int\limits_{0}^{\Lambda}\!\!d\al \: (57-70\al),
\nn\\
\rho_2^{\GG \qq^2}(s) &=& {\Gd \qq^2 \over 3^3 \cdot 2^{13} \:\pi^4} \Bigg\{
  \int\limits_{0}^{\Lambda}\!\!d\al \: (91 - 61\al) -
  \int\limits_{0}^{1}\!\!d\al \:{16m_c^2\al^2 \over (1-\al)^2} \delta \!\left( \!s \!-\! {m_c^2 \over 1\!-\!\al} \right)
  \Bigg\}
\nn
\label{dim8}
\enqa
where the integration limit is given by $\Lambda={1-m_c^2/s}$. We also have used the
definition ${\cal H}_\al = m_c^2 - (1-\al)s$, and $j=1,2$ for the currents $\eta_{1c}$
and $\eta_{2c}$, respectively.

For the $1$-structure, we get:
\beqa\label{1stru}
&&\rho_1^{pert}(s)=0,
\nn\\
&&\rho_1^{\qq}(s)=(-1)^{j+1} {\qq \over 3^2 \cdot 2^{11} \:\pi^6}\int\limits_{0}^{\Lambda}
\!\!d\al \:{\al^3 \:{\cal H}_\al^4\over(1-\al)^3} ,
\nn\\
&&\rho_1^{\GG}(s)=0,
\nn\\
&&\rho_1^{\qGq}(s)=(-1)^{j+1} {\qGq \over 3\cdot2^{11} \:\pi^6}\int\limits_{0}^{\Lambda}
\!\!d\al \:{\al^2 \:{\cal H}_\al^3 \over (1-\alpha)^2} ,
\nn\\
&&\rho_1^{\qq^2}(s)=-{m_c\qq^2 \over 3\cdot 2^7 \:\pi^4}\int\limits_{0}^{\Lambda}
\!\!d\al \:{\al^2 \:{\cal H}_\al^2\over(1-\alpha)^2} ,
\nn\\
&&\rho_1^{\GGG}(s)=0,
\nn\\
&&\rho_1^{\qq\GG}(s)=(-1)^{j+1} {\qq \Gd \over 3^3 \cdot 2^{17} \:\pi^6}\int\limits_{0}^{\Lambda}
\!\!d\al \:{\al \:{\cal H}_\al\over(1-\al)^3} \Big[ 64m_c^2 \al^2 + 3{\cal H}_\al (1-\al) (142-85\al) \Big],
\nn\\
&&\rho_1^{\qq \qGq}(s)= - {m_c \qq \qGq \over 3 \cdot 2^{11} \:\pi^4} \int\limits_{0}^{\Lambda}
\!\!d\al \:{\al \:{\cal H}_\al\over(1-\al)^2} (35 - 43\al),
\nn\\
&&\rho_1^{\qq^3}(s)= (-1)^j {\qq^3 \over 3^2 \cdot 2^{3} \:\pi^2	} \int\limits_{0}^{\Lambda}
\!\!d\al \:{\cal H}_\al ,
\nn\\
&&\rho_1^{\GG \qGq}(s)= (-1)^{j+1} {\Gd \qGq \over 3^2 \cdot 2^{17} \:\pi^6	} \int\limits_{0}^{\Lambda}
\!\!d\al \:{1 \over (1-\al)^2} \Big[ 16m_c^2 \al^2 +3{\cal H}_\al (1-\al)(13+6\al) \Big] ,
\nn\\
&&\rho_1^{\qq \GGG}(s)= (-1)^{j+1} {\qq \G3 \over 3^3 \cdot 2^{15} \:\pi^6} \int\limits_{0}^{\Lambda}
\!\!d\al \:{\al^2 \over (1-\al)^3} \bigg[ 2m_c^2(57-53\al) +{\cal H}_\al (171-167\al) \bigg],
\nn\\
&&\rho_1^{\qGq^2}(s)= -{m_c \qGq^2 \over 3 \cdot 2^{13} \:\pi^4} \int\limits_{0}^{\Lambda}
\!\!d\al \: \bigg({19 - 35\al \over 1-\al} \bigg),
\nn\\
&&\rho_1^{\GG \qq^2}(s)= -{m_c \Gd \qq^2 \over 3^3 \cdot 2^{12} \:\pi^4} \Bigg\{
  \int\limits_{0}^{\Lambda} \!\!d\al \: \bigg({65 - \al(193-152\al) \over (1-\al)^2} \bigg) -
  \int\limits_{0}^{1} \!\!d\al \:{8m_c^2 \al^2 \over (1-\al)^3} \delta \!\left( \!s \!-\! {m_c^2 \over 1 \!-\! \al} \right)
  \Bigg\} \nn.
\enqa



\begin{references}

\bibitem{Lees:2012kc}
  J.P. Lees {\it et al.}  [BaBar Collaboration],
  Phys. Rev. D {\bf 86}, 091102 (2012)
  [arXiv:1208.3086].

\bibitem{He:2012zd}
  J. He, D.-Y. Chen and X. Liu,
  Eur. Phys. J. C {\bf 72}, 2121 (2012) 
  [arXiv:1204.6390].

\bibitem{Zhang:2012xx}
  J.-R. Zhang,
  Phys. Rev. D {\bf 87}, 076008 (2013)
  [arXiv:1211.2277].

\bibitem{svz} 
  M.A. Shifman, A.I. Vainshtein and V.I. Zakharov,
  Nucl. Phys. {\bf B147}, 385 (1979).

\bibitem{rry} 
  L.J. Reinders, H. Rubinstein and S. Yazaki,
  Phys. Rept. {\bf 127}, 1 (1985).

\bibitem{SNB} 
  For a review and references to original works, see e.g., 
  S. Narison, {``QCD as a Theory of Hadrons'', Cambridge 
  Monogr. Part. Phys. Nucl. Phys. Cosmol.} {\bf 17}, 1 (2002) [hep-h/0205006]; 
  {\it ibid.}, {``QCD spectral sum rules'', World Sci. Lect. Notes Phys.} {\bf 26}, 1 (1989); 
  {\it ibid.}, {Acta Phys. Pol.} {\bf B26}, 687 (1995); 
  {\it ibid.}, {Riv. Nuov. Cim.} {\bf 10N2}, 1 (1987); 
  {\it ibid.}, {Phys. Rept.} {\bf 84}, 263 (1982).

\bibitem{Diakonov:2010tf}
  D. Diakonov,  arXiv:1003.2157.

\bibitem{Wu:2004wg} 
  B. Wu and B.-Q. Ma,
  Phys. Rev. D {\bf 70}, 034025 (2004)
  [hep-ph/0402244].

\bibitem{Kim:2004pu}
  H. Kim, S.H. Lee and Y. Oh,
  Phys. Lett. B {\bf 595}, 293 (2004)
  [hep-ph/0404170].

\bibitem{Shuryak:2005pk}
  E.V. Shuryak,
  J. Phys. Conf. Ser. {\bf 9}, 213 (2005) 
  [hep-ph/0505011].

\bibitem{Lee:2009rt}
  S. H. Lee and S. Yasui,
  Eur. Phys. J. C {\bf 64}, 283 (2009)
  [arXiv:0901.2977].  

\bibitem{narpdg} 
  S. Narison, Phys. Lett. {\bf B624}, 223 (2005); 
  {\it ibid.}, Phys. Lett. {\bf B466}, 345 (1999);
  {\it ibid.}, Phys. Lett. {\bf B387}, 162 (1996);
  {\it ibid.}, Phys. Lett. {\bf B361}, 121 (1995);
  H. Forkel and M. Nielsen, Phys. Lett. {\bf B345}, 55 (1995).

\bibitem{nfl}
  M. Nielsen, F.S. Navarra and S.H. Lee,
  Phys. Rept. {\bf 497}, 41 (2010)
  [arXiv:0911.1958].

\end{references}
\end{document}